\documentclass[%
reprint,
superscriptaddress,
 amsmath,amssymb,
 aps,
pra,
]{revtex4-2}
\usepackage{graphicx}
\usepackage{dcolumn}
\usepackage{bm}
\usepackage{comment}
\usepackage{braket}
\usepackage{float}
\usepackage[caption=false]{subfig}
\usepackage{array}
\newcolumntype{P}[1]{>{\centering\arraybackslash}p{#1}}
\newcolumntype{M}[1]{>{\centering\arraybackslash}m{#1}}
\newcommand{\rvline}{\hspace*{-\arraycolsep}\vline\hspace*{-\arraycolsep}}

\begin{document}

\preprint{APS/123-QED}

\title{Higher-dimensional Hong-Ou-Mandel effect and state redistribution with linear-optical multiports
}

\author{Shuto Osawa}
\email[e-mail: ]{sosawa@bu.edu} \affiliation{Dept. of Electrical and Computer Engineering \& Photonics Center, Boston University, 8 Saint Mary's St., Boston, MA
02215, USA}
\author{David S. Simon}
\email[e-mail: ]{simond@bu.edu} \affiliation{Dept. of Physics and Astronomy, Stonehill College, 320 Washington Street, Easton, MA 02357} \affiliation{Dept. of
Electrical and Computer Engineering \& Photonics Center, Boston University, 8 Saint Mary's St., Boston, MA 02215, USA}
\author{Alexander V. Sergienko}
\email[e-mail: ]{alexserg@bu.edu} \affiliation{Dept. of Electrical and Computer Engineering \& Photonics Center, Boston University, 8 Saint Mary's St., Boston,
MA 02215, USA} \affiliation{Dept. of Physics, Boston University, 590 Commonwealth Ave., Boston, MA 02215, USA}

\date{\today}

\begin{abstract}
We expand the two-photon Hong-Ou-Mandel (HOM) effect onto a higher dimensional set of spatial modes and introduce an effect that allows controllable redistribution of quantum states over these modes using directionally-unbiased linear-optical four-ports without post-selection. The original HOM effect only allows photon pairs to exit in two directions in space.
But when accompanied by beam splitters and phase shifters, the result is a directionally-controllable two-photon HOM effect in four spatial modes, with direction controlled by changing the phases in the system.
This controllable quantum amplitude manipulation also allows demonstration of a “delayed” HOM effect by exploiting phase shifters in a system of two connected multiport devices.
By this means, both spatial and temporal control of the propagation of the two-photon superposition state through a network can be achieved.

\end{abstract}

\maketitle



\begin{figure}
\vspace{-5mm}
	\centering
	\includegraphics[width=3in]{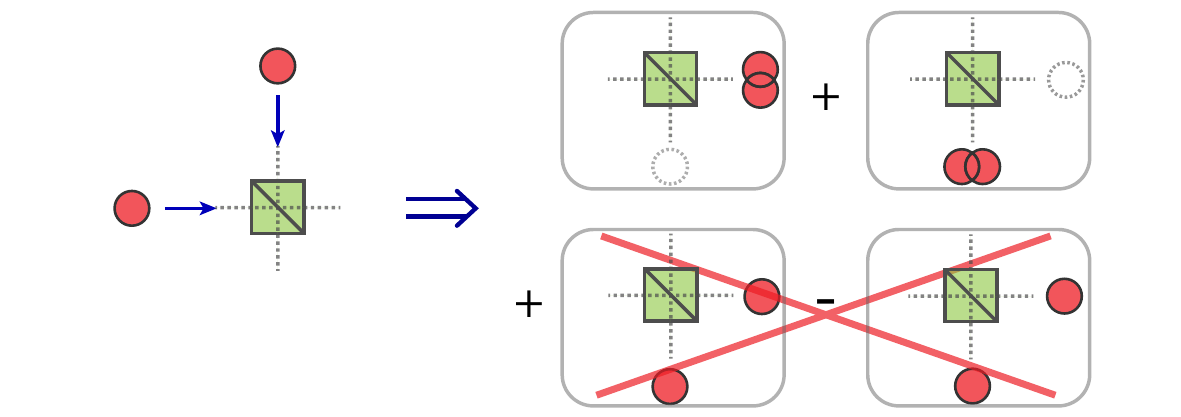}
	\caption{Hong-Ou-Mandel effect. Two identical photons are sent into different beam splitter input ports. There are four possible outcomes, two photons leaving one port, two photons leaving the other port, each photon reflecting to give single photons at each exit, and both transmitting to give single photons at each port. The coincidence terms cancel out since they are identical but enter with opposite sign. The final state is a superposition of two outcomes, each with both photons clustered together at the same exit port.}
	\label{fig:HOM_effect}
\end{figure}

\section{Introduction}
\vspace{-10px}
The Hong-Ou-Mandel (HOM) effect is one of the most recognized quantum two-photon interference effects \cite{hong1987measurement}.
When two indistinguishable photons arrive simultaneously at different inputs of a 50:50 beam splitter (BS), single-photon amplitudes at each output cancel, resulting in quantum superposition of two-photon states appearing at each output port, as in Fig. \ref{fig:HOM_effect}.
This traditional HOM method, observed on a BS having two input and two output ports, always has the two-photon state simultaneously occupying both output spatial modes, leaving no room to engineer control of propagation direction.

Various types of studies on quantum state transformations in multiport devices have been performed such as two photon propagation in a multimode system \cite{weihs1996two,zukowski1997realizable},
quantum interference effects using a few photons \cite{meany2012non,de2014coincidence,tichy2011four,campos2000three}, and propagation of multi-photons \cite{lim2005generalized,tillmann2015generalized,menssen2017distinguishability}. Internal degrees of freedom are also incorporated to enhance communication capacity \cite{walborn2003multimode,poem2012two,zhang2016engineering}.
Systems and procedures using multi-photon states, such as boson sampling, have been analyzed using multiport beam splitters both theoretically and experimentally \cite{aaronson2011computational,tillmann2013experimental,spring2013boson,bentivegna2015experimental,he2017time,wang2019boson}.
The HOM effect plays an important role in the field of quantum metrology when two-photon $|2002\rangle$-type states are extended to $N$-photon $N00N$ state \cite{dowling2008quantum,motes2015linear}.

Additionally, coherent transport of quantum states has been attracting attention, where single- and two-photon discrete-time quantum walk schemes are employed to transfer and process quantum states \cite{bose2003quantum,perez2013coherent,lovett2010universal,chapman2016experimental,nitsche2016quantum}.
A quantum routing approach has been proposed to transfer unknown states in 1D and 2D structures to assist quantum communication protocols \cite{zhan2014perfect,vstefavnak2016perfect,bartkiewicz2018implementation}.

Photon propagation control is especially crucial in a large optical network to distribute quantum states between two parties.
The network can be formed by combining multiple copies of four-port devices.
The state manipulation schemes we present can be integrated in quantum communication protocols since state retrieval timing can be chosen at will.

In this manuscript, we propose two-photon quantum state engineering and transportation methods with a linear-optical system which allows manipulation of photon amplitudes by using linear-optical devices such as optical multiports, beam splitters, and phase shifters.

Previously, such multiports have been introduced to demonstrate a two-photon clustering effect in quantum walks when multiple multiport devices are connected to form a chain \cite{simon2020quantum}. Clustering of two photons means that after encountering a multiport, the input two-photon amplitude separates into a superposition of a right-moving and a left-moving two-photon amplitude, with no amplitude for the photons to move in opposite directions.
By utilizing this separation, a higher-dimensional unitary transformation enables flexible quantum engineering designs of possible travel path combinations by switching relative phases within right moving and left moving amplitudes independently.
When two or more multiports are combined, this control of quantum amplitudes in a two-photon state allows demonstration of a “delayed" HOM effect engaging also time-bin modes in addition to spatial modes.
To perform this delayed effect, two or more multiports are required, and relative phase shifts between two rails can reflect the incoming amplitudes.
This controllable reflection without mirrors can also be seen as an additional state manipulation feature.

We introduce two distinct systems. The first system utilizes direct transformation of two-photons by the four-port device using circulators.
The second case does not have circulators in the system.
The photons are sent from the left side of the beam splitters, then the amplitudes encounter the multiport device.
This second system has not been analyzed in the past. Specific input states for both distinguishable and indistinguishable photons are redistributed between two parties coherently. Therefore, this system is particularly useful in quantum routing type applications.

This article is organized as follows. In Sec. II, we introduce the main optical components used in this manuscript to perform quantum state transformation. These basic linear optical devices are used to show HOM effect engaging in spatial modes and time-bin modes, and is addressed in Sec. III. In Sec. IV, we show redistribution of two photon states using the devices introduced that are presented in Sec. II. The summary of the results are given in Sec. V.

\begin{figure}[htp]
\subfloat[]{%
  \includegraphics[clip,width=0.8\columnwidth]{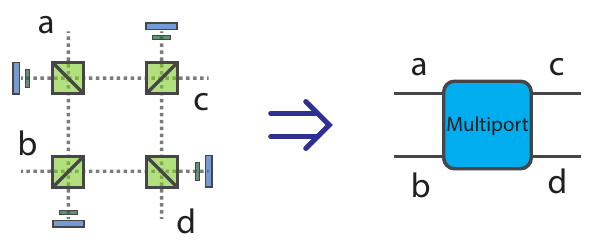}%
}
\vspace{-10px}
\subfloat[]{%
  \includegraphics[clip,width=0.8\columnwidth]{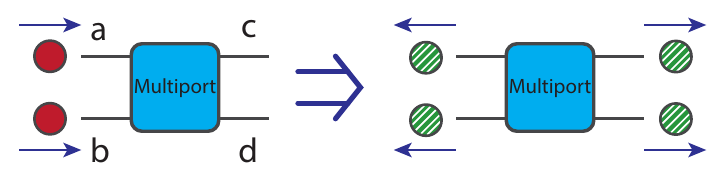}%
}
\vspace{-10px}
\caption{(a) A possible experimental realization of a directionally-unbiased linear-optical four-port consists of four beam splitters, four mirrors, and four phase shifters. A photon can enter any of the four ports, and exit at any of the four ports (labeled as $a,b,c$, and $d$). With a specific choice of phase settings, a Grover matrix can be realized by coherently summing all possible paths to each output \cite{osawa2019directionally}. A schematic symbol for this device is shown on the right. (b) Single multiport transformation of a two-photon input state. The input state of two correlated photons entering from the left is depicted as $ab$ $(\ket{1,1})$. After scattering by a Grover multiport, the state transforms into $-\frac{1}{4}(a-b)^2+\frac{1}{4}(c+d)^2$, which has clear separation of right- and left-moving two-photon amplitudes. No cross-terms with photons moving in opposite directions occur.}
\label{fig:combined_four_port}
\vspace{-10pt}
\end{figure}

\section{Photonic state transformations via linear optical devices}
\vspace{-10px}
In this section, we consider photonic state transformations in higher-dimensional spatial modes using a unitary four-dimensional Grover matrix \cite{grover1996fast} in place of the beam splitter.
In this section, we introduce the main systems that will be used for linear state transformations, followed by the basic photonic devices to implement them.
Beam splitters and the four-dimensional Grover matrix are the central system component.
We mainly use photon number representation to describe states through out the manuscript.
The general beam splitter transformation matrix is
\begin{eqnarray}
\label{eqn:BS}
&\begin{pmatrix}
\hat{c}
\\
\hat{d}
\end{pmatrix}=
\frac{1}{\sqrt{2}}
\begin{pmatrix}
1 & 1\\
-1 & 1
\end{pmatrix}
\begin{pmatrix}
\hat{a}\\
\hat{b}
\end{pmatrix}
\end{eqnarray}
where $\hat{a}$,$\hat{b}$,$\hat{c}$, and $\hat{d}$ are used to describe the input photon state transformation.
The labels are generalized here, therefore the specific location dependent beam splitter transformations are redefined in later sections.
We use photon number states to describe the system unless otherwise specified. The input state is denoted as $\hat{a}\hat{b}$ where $\hat{a}$ and $\hat{b}$ are respectively creation operators for the spatial modes $a$ and $b$.
The hat notation is dropped henceforth.
For a photon in spatial mode $a$ with horizontally polarized photon is denoted as $a_{H}$ and horizontally polarized photon in mode $b$ is denoted as $b_{H}$.
We omit polarization degrees of freedom when identical photons are used through out the system.
Photonic implementations of the Grover matrix can be readily realized \cite{carolan2015universal,crespi2013anderson,spagnolo2013three,fan1998channel,nikolopoulos2008directional}.
To be concrete, we use directionally-unbiased linear-optical four-ports (Fig. \ref{fig:combined_four_port} (a)) as an example.
Consider sending two indistinguishable photons into a four-dimensional multiport device realization of a Grover matrix.
This Grover operator, the multiport, described by the unitary matrix
\begin{equation}
\label{eqn:Grover}
Grover =
\frac{1}{2}
\begin{pmatrix}
-1&1&1&1\\
1&-1&1&1\\
1&1&-1&1\\
1&1&1&-1
\end{pmatrix},
\end{equation}
has equal splitting ratios between all input-output combinations and generalizes the BS transformation matrix given below in Eq.\eqref{eqn:BS}.
In general, photons in modes $a$ and $b$ are transformed in the following manner,
\begin{eqnarray}
&a \xrightarrow{M} \frac{1}{2}(-a+b+c+d)\; \mbox{ and }\;\\ \nonumber
&b \xrightarrow{M} \frac{1}{2}(a-b+c+d).
\end{eqnarray}
Theoretical analysis of the reversible Grover matrix has been performed by linear-optical directionally-unbiased four-ports  \cite{simon2016group,simon2018joint,osawa2019directionally}, which consist of four beam splitters, four phase shifters, and four mirrors as indicated in Fig. \ref{fig:combined_four_port}(a).
They are represented schematically by the symbol in Fig. \ref{fig:combined_four_port}(b).
The three-port version of this device has been experimentally demonstrated using bulk optical devices \cite{osawa2018experimental}.
To have better and precise control of phases, miniaturization of the device is highly preferred to realize the four-port especially when several multiport devices are required to carry out an experiment.
In general, directional unitary devices such as those of the Reck and some other unitary matrix decomposition models \cite{reck1994experimental,su2019hybrid,clements2016optimal,de2018simple,motes2014scalable} can also realize a Grover matrix.
However, directionally-unbiased devices are advantageous when designing the delayed HOM effect, as well as requiring fewer optical resources.
Identical photons are sent into two of the four input-output ports from the left side (indicated in Fig. \ref{fig:combined_four_port}(b)).
We used multiport devices and beam splitters to form two systems for state propagation.
The photons are sent from the left side of the system through out the manuscript.
The first BS multiport composite system is denoted as subscript 0 and the other half is denoted as subscript 1.
The result differs depending on the input location of photons.
Consider a system consisting of two multiports and two beam splitters.
There are several ways to insert photons in the system, however we choose two specific ones in this manuscript.
To be able to send a photon into the middle of the system, the setup needs to be supplied with circulators, is shown in Fig. \ref{fig:multiport_circ}.
Another setup requires no circulators to propagate input photons.
The photons experience an extra transformation by a beam splitter upon photon entrance.
The system is graphically supplied in Fig. \ref{fig:multiport_no_circ}.
It needs to be noted that the number of multiports in the system does not change the final outcome.
We are using two multiports as an example, however, the result is the same when the system has a single multiport or more than two multiports as long as the devices are assumed to be lossless during the propagation.

Brief comments on the mathematical structure of the transformations carried out by the configurations given in Figs. \ref{fig:multiport_circ} and \ref{fig:multiport_no_circ} are given in the Appendix.

\vspace{-15px}
\subsection{Photon propagation using circulators}
\vspace{-10px}
\begin{figure}
\vspace{-5mm}
	\centering
	\includegraphics[width=3.5in]{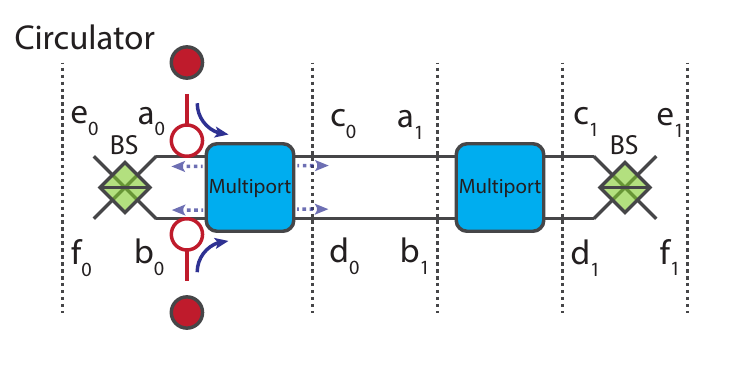}
	\caption{A system setup with input photons supplied by circulators. The system consists of two beam splitters, two multiport devices, and two circulators. These circulators allow us to send photons from the left side of the multiport device without experiencing a beam splitter transformation before entering the multiport device. The input state split into right moving and left moving amplitudes (shown as dotted arrows) upon multiport transformation.}
	\label{fig:multiport_circ}
\end{figure}
This method is used to distribute HOM pair between the right and the left side of the system.
The original input state $a_0b_0$ transforms to:
\begin{eqnarray} \label{eq:trans}
a_0b_0 &\xrightarrow{M} \frac{1}{2}(-a_0+b_0+c_0+d_0)\frac{1}{2}(a_0-b_0+c_0+d_0) \nonumber \\
&=-\frac{1}{4}(a_0^2+b_0^2)+\frac{1}{2}a_0b_0+\frac{1}{4}(c_0^2+d_0^2)+\frac{1}{2}c_0d_0\nonumber \\
&= -\frac{1}{4}(a_0-b_0)^2+\frac{1}{4}(c_0+d_0)^2,
\end{eqnarray}
where we have used the commutation relation $ab = ba$ since the photons are identical and in different spatial locations.
Eq.\eqref{eq:trans} shows that correlated photons are split into right moving $\frac{1}{4}(c_0+d_0)^2$ and left moving $-\frac{1}{4}(a_0-b_0)^2$ amplitudes, with no cross terms.
This absolute separation of propagation direction without mixing of right moving and left moving amplitudes is important because the photon pairs remain distinctly localized and clustered at each step \cite{simon2020quantum}.
The right moving amplitude is translated to $\frac{1}{4}(a_1+b_1)^2$ and propagates without changing its form.
$\frac{1}{4}(a_1+b_1)^2 \xrightarrow{M} \frac{1}{4}(c_1+d_1)^2$.
The left moving amplitude $-\frac{1}{4}(a_0-b_0)^2$ stays the same until BS transformation.


The controlled HOM effect can be observed in higher-dimensional multiports assisted by extra beam splitters.
Imagine beam splitters inserted in the system as in Fig. \ref{fig:multiport_circ}.
Input state $ab$ is now transformed into $-\frac{1}{4}(a-b)^2+\frac{1}{4}(c+d)^2$ as indicated above, then further transformed by beam splitters to obtain HOM pairs between the right side and left side of the system.
The right and left sides of the system each have two output ports, and the exit port of the photon pair can be controlled by varying phase shift settings before the beam splitters.
A phase shift on the left side of the system does not affect the result of the right side amplitude, and vice versa.
This system, having circulators at the beginning of the system, is denoted as transformation pattern I, and the detailed discussions of its transformation are in Sec. \ref{sec:pattern_I}.

\vspace{-15px}
\subsection{Photon propagation without circulators}
\vspace{-10px}
\begin{figure}
\vspace{-5mm}
	\centering
	\includegraphics[width=3.5in]{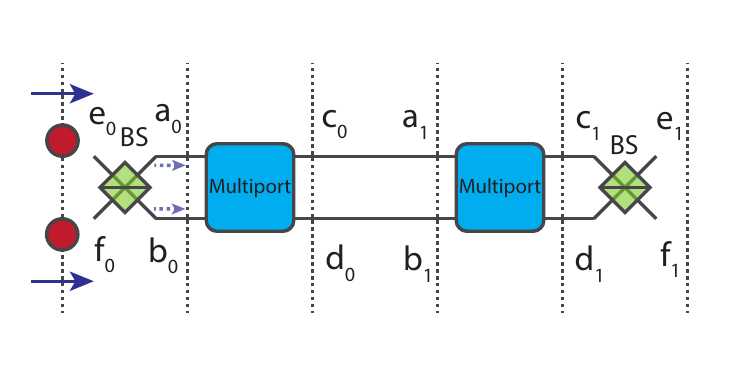}
	\caption{System setup without circulators. The input photons are subjected to a beam splitter before they enter the multiport. The input state is transformed and propagated in one direction (shown as dotted arrows). The BS transformed input state is transformed again by the first multiport devices.}
	\label{fig:multiport_no_circ}
\end{figure}
This method allows to redistribute input states between right and left side of the system without changing amplitudes.
Consider sending two photons from the left side of the beam splitter as indicated in fig. \ref{fig:multiport_no_circ} then transform the output state by the multiport device.
We only consider the first multiport transformation here.
The rest of the transformation is given in sec. \ref{sec:pattern_II}.
\begin{equation}
e_0f_0\xrightarrow{BS}-\frac{1}{2}(a_0^2-b_0^2)
\xrightarrow{M}-\frac{1}{2}(a_0-b_0)(c_0+d_0)
\end{equation}
The final state has cross-terms, and it is different from the case with circulators in a sense that the output state is \textit{coupled}.
The state does not provide clear separation between right moving and left moving amplitudes.
Even though, the state does not have clear distinction between right moving and left moving, we still refer the amplitudes right and left moving amplitudes unless special attention is required.
This system having no circulators is denoted as transformation pattern II, and the detailed discussions of its transformation are in Sec. \ref{sec:pattern_II}
\vspace{-10px}
\section{Transformation pattern I: directionally-controllable HOM effect in higher-dimensional spatial and temporal modes} \label{sec:pattern_I}
\vspace{-10px}
In this section we discuss the transformation pattern I.
The higher dimensional HOM effect is generated by the multiport-based linear optics system with circulators at the inputs.
The propagation direction control and delays between amplitudes are discussed in subsections.
We use a single multiport device to show the control effect and we introduce two multiport devices in the system for delayed effect.

\begin{figure}[htp]
\subfloat[]{%
  \includegraphics[clip,width=\columnwidth]{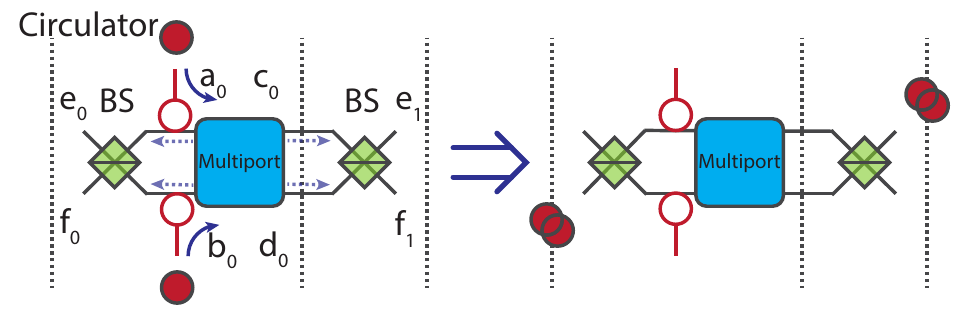}%
}
\hspace{-10px}
\subfloat[]{%
  \includegraphics[clip,width=\columnwidth]{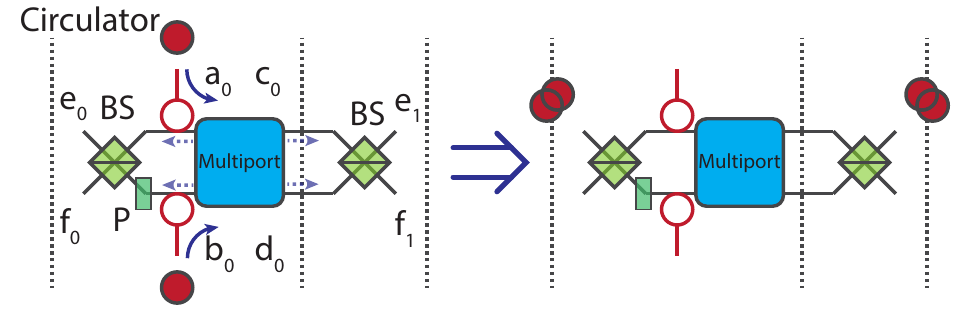}%
}
\hspace{-10px}
\subfloat[]{%
  \includegraphics[clip,width=\columnwidth]{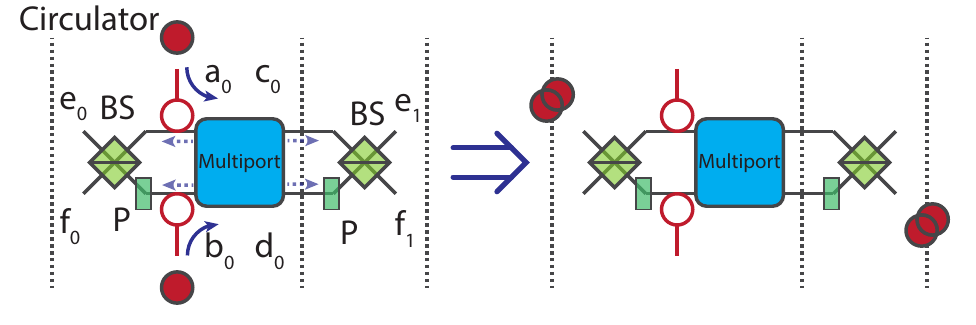}%
}
\hspace{-10px}
\subfloat[]{%
  \includegraphics[clip,width=\columnwidth]{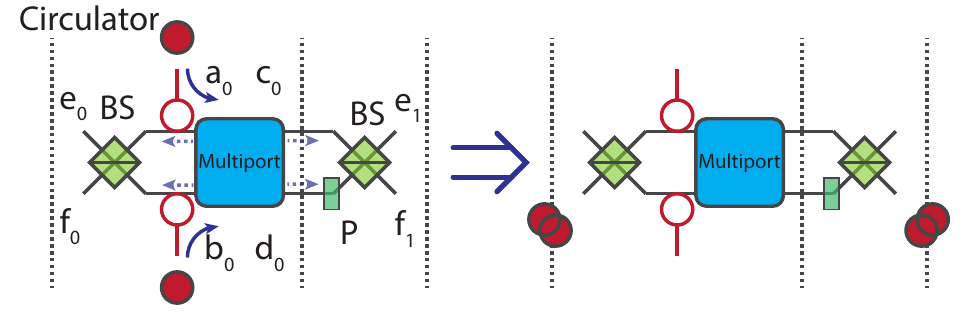}%
}

\caption{Higher dimensional HOM effect with directional control.
Correlated photons, $a_0b_0$, are sent in from the circulators into the first multiport.
After the first multiport interaction, the incoming photon pair splits into right-moving and left-moving two-photon amplitudes.
The separately-moving amplitudes are bunched at the beam splitters on right and left sides.
We can controllably switch between four different output sites, and where the clustered output photons appear depends on the location of the phase shifter $P$.
In (a), no phase plates are introduced, and the output biphoton amplitudes leave $f_0$ and $e_1$. The final state is $\frac{1}{\sqrt{2}}(\ket{0,2}_0+\ket{2,0}_1)$, meaning superposition of two photons in mode $f_0$ and two in mode $e_1$. In case (b), the phase shifter $P=\pi$ is to the left, changing the relative phase between upper and lower arms.
Similarly in (c) and (d), other locations for the phase shifters cause biphotons to leave in other spatial modes.} \label{fig:HOM_control_fig}
\end{figure}
\vspace{-15px}
\subsection{Control of propagation direction}
\vspace{-10px}
Given that one two-photon amplitude must exit left and one right, there are four possible combinations of outgoing HOM pairs as indicated in Fig. \ref{fig:HOM_control_fig}.
The combinations are, (a): $(f_0^2,e_1^2)$, (b): $(e_0^2,e_1^2)$, (c): $(e_0^2,f_1^2)$, and (d): $(f_0^2,e_1^2)$.
This means, in the case of (a) for example, the left-moving two-photon amplitude leaves in mode f, and the right-moving amplitude leaves in mode e.
Directional control of the four cases is readily demonstrated, as follows. In case (a) there is only a beam splitter transformation after the multiport, giving
\begin{eqnarray}
&-\frac{1}{4}(a_0-b_0)^2\xrightarrow{BS} -\frac{1}{4\sqrt{2}}(e_0-f_0-e_0-f_0)^2 = \frac{1}{2}f_0^2, \nonumber \\
&\frac{1}{4}(c_1+d_1)^2 \xrightarrow{BS} -\frac{1}{4\sqrt{2}}(e_1-f_1+e_1+f_1)^2 = \frac{1}{2}e_1^2.
\end{eqnarray}
The final output state is,
\begin{eqnarray}
\frac{1}{2}(f_0^2+e_1^2)=\frac{1}{\sqrt{2}}(\ket{0,2}_0+\ket{2,0}_1).
\end{eqnarray}
In case (b), a phase plate is inserted in the lower arm of the left side to switch the exit port from $d$ to $c$. All the phase shifters P are set to $\pi$, therefore transforming $b \rightarrow -b$.
\begin{eqnarray}
&-\frac{1}{4}(a-b)^2+\frac{1}{4}(c+d)^2 \xrightarrow{P} -\frac{1}{4}(a+b)^2+\frac{1}{4}(c+d)^2 \nonumber \\
&\xrightarrow{BS} \frac{1}{2}(-e_0^2+e_1^2) = \frac{1}{\sqrt{2}}(-\ket{2,0}_0+\ket{2,0}_1).
\end{eqnarray}
Compared to case (a), the exit port is switched from f to e. In (c), phase plates are inserted in the lower arms of both right and left sides.
Photons in modes $b$ and $d$ are transformed to $-b$ and $-d$, respectively.
\begin{eqnarray}
&-\frac{1}{4}(a-b)^2+\frac{1}{4}(c+d)^2 \xrightarrow{P} -\frac{1}{4}(a+b)^2+\frac{1}{4}(c-d)^2 \nonumber \\
&\xrightarrow{BS} \frac{1}{2}(-e_0^2-f_1^2) = -\frac{1}{\sqrt{2}}(\ket{2,0}_0+\ket{0,2}_1).
\end{eqnarray}

In (d), a phase plate is inserted in the lower arm of the right side. A photon in mode $d$ is transformed to $-d$.
\begin{eqnarray}
&-\frac{1}{4}(a-b)^2+\frac{1}{4}(c+d)^2 \xrightarrow{P} -\frac{1}{4}(a-b)^2+\frac{1}{4}(c-d)^2 \nonumber \\
&\xrightarrow{BS} \frac{1}{2}(f_0^2-f_1^2) = \frac{1}{\sqrt{2}}(\ket{0,2}_0-\ket{0,2}_1).
\end{eqnarray}
This demonstrates complete directional control of biphoton propagation direction using only linear optical devices.
Directional control does not require changing splitting ratios at each linear optical device (BS and multiport), and occurs in a lossless manner since no post-selection is required.

\begin{figure*}[htp]
\centering
\subfloat[][Delayed HOM effect without reflection]{\includegraphics[clip,width=\columnwidth]{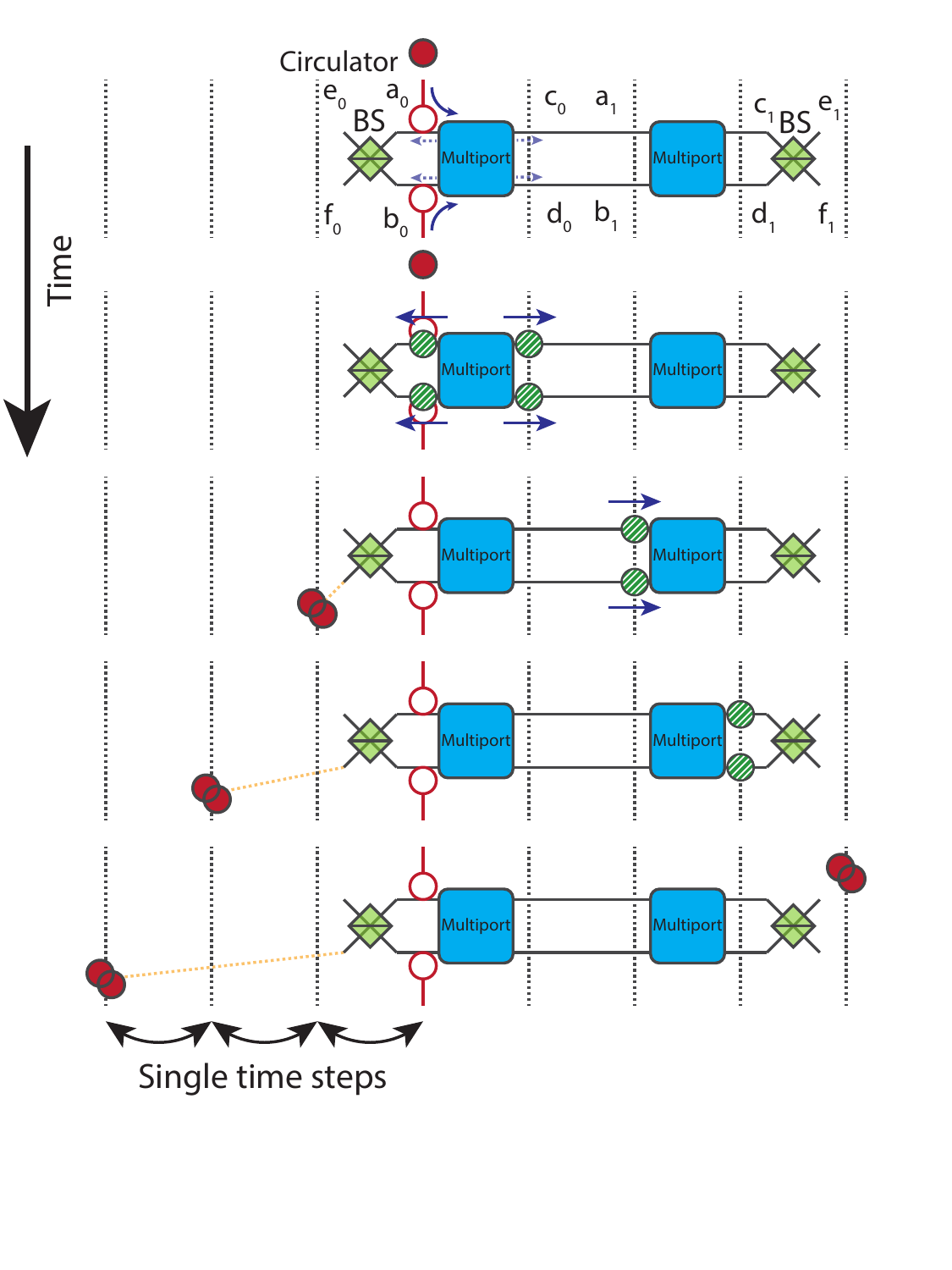}\label{}}
\subfloat[][Delayed HOM effect with reflection]{\includegraphics[clip,width=\columnwidth]{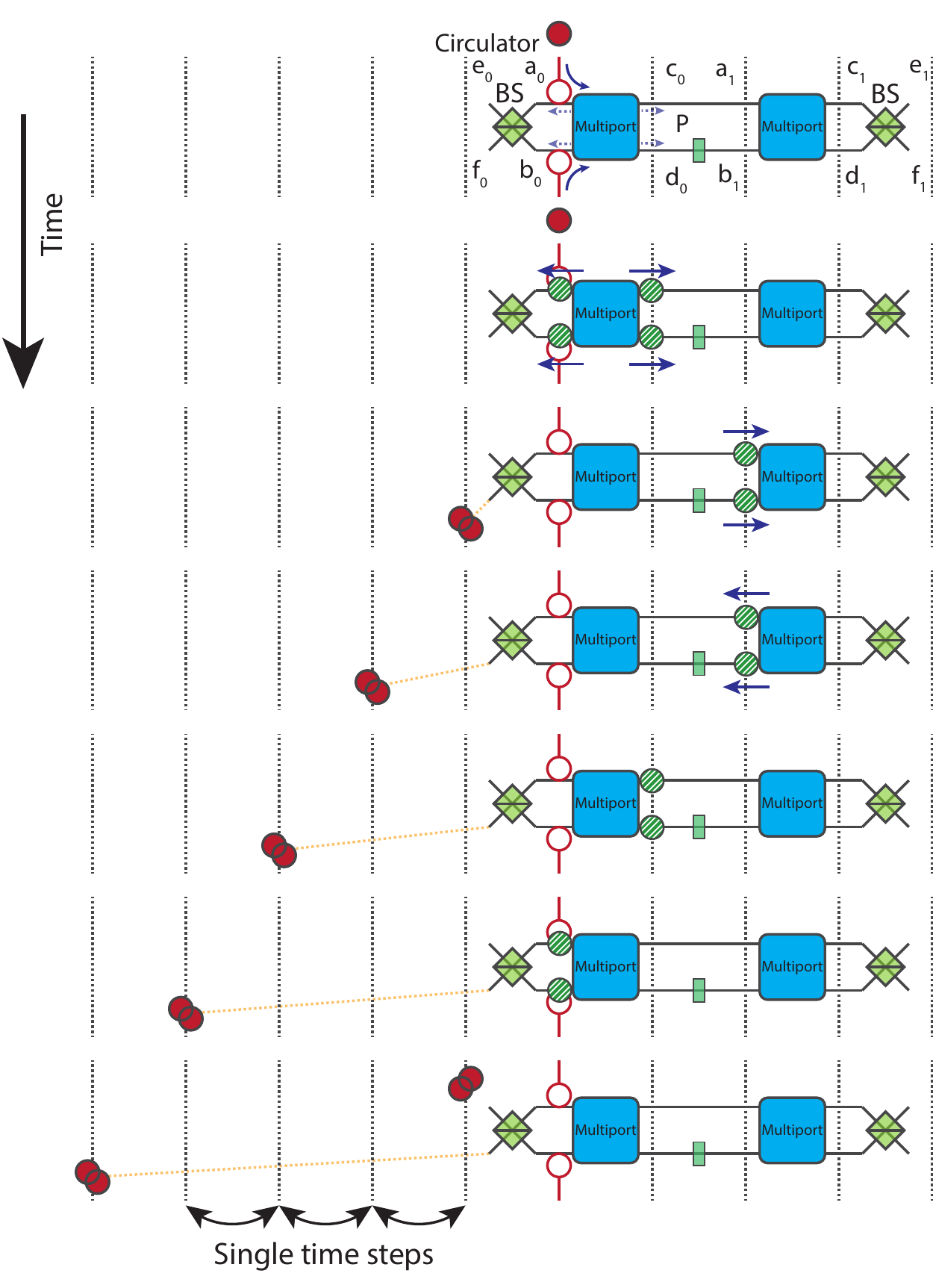}\label{}}
\caption{Delayed HOM effect. The two-photon amplitude transformation progresses in time from top to bottom. The distance traveled in a single time step is indicated by vertical dashed lines. The original photons as well as photons in the target state are indicated using red circles. The green striped circles indicate intermediate transformed state. The total number of photons are always two through out the transformations. (a) Two multiports and beam splitters {\it without} phase shifters between the multiports. At the first step, the behavior is the same as for a single multiport with beam splitters. The right-moving amplitude propagates through the second multiport, and left-moving amplitude propagates through the beam splitter. The right moving amplitude is delayed by one additional multiport transformation before a two-photon observation probability will become available in spatial modes on the right. (b) Two multiports and beam splitters {\it with} a phase shifter P set at $\pi$ between multiports. When the P is present, the right-moving amplitude gains a relative phase between modes $
a_1$ and $b_1$. Reflection occurs at the multiport when the relative phase between the two is $\pi$. Therefore, the transformed amplitude  reflects upon a second multiport encounter, going back to the original state with opposite propagation direction. Reflection does not occur on this transformed left-moving amplitude, therefore it continues to propagate leftward. The original left-moving amplitude becomes available for detection earlier than the transformed left-moving amplitude.}
\label{fig:delayed_HOM}
\end{figure*}


\subsection{Delayed HOM effect}
\vspace{-10px}
\subsubsection{Delayed HOM effect without reflection}
\vspace{-10px}
We introduce a phase shifter between two multiports as in Fig. \ref{fig:delayed_HOM} (b). Without the phase plate between two multiport devices, the photons behave exactly the same as in the previous subsection. However, the phase shifter can change propagation direction of right moving amplitude to the left. This reflection results in detecting HOM pairs only on the left side, but with some delay between the two exiting amplitudes. We start with the case without the phase shifter. The photon insertion is the same as the previous case, coming from the left side of the first multiport.
\vspace{-10px}
\begin{eqnarray}
    &{a_{0}b_{0}}_R\xrightarrow{M}-\frac{1}{4}(a_{0}-b_{0})_L^2+ \frac{1}{4}(c_{0}+d_{0})_R^2 \nonumber \\
    &\xrightarrow{T+BS,\mbox{ }T} -\frac{1}{2}f_{0L}^2 + \frac{1}{4}(a_{1}+b_{1})_R^2  \nonumber \\
    &\xrightarrow{M} -\frac{1}{2}f_{0L}^2 + \frac{1}{4}(c_{1}+d_{1})_R^2 \xrightarrow{BS} -\frac{1}{2}f_{0L}^2 + \frac{1}{2}e_{1R}^2,
\end{eqnarray}
where M, T, BS represents multiport, translation and beam splitter transformation respectively. We use subscript $R$ and $L$ to illustrate amplitudes propagating to the right or left. $T$ translates a photon amplitude by a single time step (for example, $\frac{1}{4}(c_{0}+d_{0})^2 \rightarrow \frac{1}{4}(a_{1}+b_{1})^2$). The second transformation $T+BS$, $T$ is read as applying $T+BS$ on the first term and $T$ on the second term.

The final state is,
\begin{eqnarray}
    -\frac{1}{2}f_{0L}^2 + \frac{1}{2}e_{1R}^2=-\frac{1}{\sqrt{2}}(\ket{0,2}_{0{T_0}L}-\ket{2,0}_{1{T_1}R}),
\end{eqnarray}
where $T_0$ is the time when the first biphoton amplitude leaves the system and $T_1$ is the exit time of the second.
The right moving amplitude stays in the system longer than the left moving amplitude because of the extra multiport device in the system, leading to time delay $\Delta T = T_1-T_0$.
\vspace{-10px}
\subsubsection{Delayed HOM effect with reflection}
\vspace{-10px}
When a $\pi$-phase shifter is inserted on one path between the multiports, the right-moving amplitude gets reflected upon the second multiport encounter.
Instead of having two-photon amplitudes on the right and left sides of the system, both photon amplitudes end up leaving from the left.
The HOM effect still occurs but now with some delay between the two amplitudes at the end of the BS.
This is indicated in Fig. \ref{fig:delayed_HOM} (b).
\vspace{-10px}
\begin{eqnarray}
    &{a_{0}b_{0}}_R\xrightarrow{M}-\frac{1}{4}(a_{0}-b_{0})_L^2+ \frac{1}{4}(c_{0}+d_{0})_R^2 \nonumber \\
    &\xrightarrow{T+BS,\mbox{ }T+P} -\frac{1}{2}f_{0L}^2 + \frac{1}{4}(a_{1}-b_{1})_R^2.
\end{eqnarray}
The second transformation $T+BS$, $T+P$ is read as applying $T+BS$ on the first term and $T+P$ on the second term.
Left-moving photons leave before right-moving photons.
\begin{eqnarray}
    \xrightarrow{M}&\frac{1}{4}(a_{1}-b_{1})_L^2 \xrightarrow{P+T} \frac{1}{4}(c_{0}+d_{0})_L^2 \nonumber \\
    &\xrightarrow{M} \frac{1}{4}(a_{0}+b_{0})_L^2 \xrightarrow{BS} \frac{1}{2}e_{0L}^2.
\end{eqnarray}
The final state,
\begin{eqnarray}
    -\frac{1}{2}f_{0L}^2+\frac{1}{2}e_{0L}^2=-\frac{1}{\sqrt{2}}(\ket{0,2}_{0T_0L}-\ket{2,0}_{0T_2L}),
\end{eqnarray}
is now two HOM pair amplitudes, both on the left side of the system, at output ports $e_0$ and $f_0$, with some time delay $\Delta T = T_2 - T_0$ between them. The first amplitude leaves port $f_0$ at $T_0$, then the second leaves $e_0$ and the time labeled $T_2$.
\vspace{-10px}
\section{Transformation pattern II: state redistribution in higher-dimensional spatial and temporal modes} \label{sec:pattern_II}
\vspace{-10px}
\subsection{State transformation and propagation}
\vspace{-10px}
We have considered the case where the input photon state is transformed by the multiport device right after photon insertion in the previous section.
Instead of using circulators, we can transform the input state by the BS in advance and then transform the state by using the multiport device.
Even though the Grover matrix spreads the input state equally in four directions, the end result preserves the original form of the input state.
We demonstrate a state redistribution property using distinguishable and indistinguishable photons, meaning the input state gets redistributed between right and left side without changing amplitudes.
The propagation result is different from the previous case.
Consider sending two indistinguishable photons in the system.
The input two photons have the same polarization to make them indistinguishable.
The input photons are inserted from the left side of the beam splitter.
The beam splitter transforms the input state and propagates from the left side to right side of the device without any reflections.
The amplitudes are transformed by the multiport device after the beam splitter transformation.
This transformation splits input photons into coupled right- moving and left-moving amplitudes.
The coupled left moving amplitudes reflected from the first multiport counter propagates and transformed by the first beam splitter from the right to the left.
The right moving amplitude is transmitted without changes in amplitude.
This amplitude gets transmitted by the right side beam splitter at the end.
\vspace{-15px}
\subsubsection{Indistinguishable photons}
\vspace{-10px}
We examine the mathematical details on indistinguishable photons in the system without circulators first.
We consider three cases by sending photons in spatial modes e and f.
First, we consider indistinguishable a pair of single photons from spatial mode e and f.
\vspace{-5px}
\begin{eqnarray}
&e_{H0}f_{H0}\xrightarrow{BS}-\frac{1}{2}(a_{H0}^2-b_{H0}^2)\nonumber \\
&\xrightarrow{M} -\frac{1}{2}(a_{H0}-b_{H0})(c_{H0}+d_{H0}) \nonumber \\
&\xrightarrow{BS} -e_{H0}e_{H1}.
\end{eqnarray}

HOM state with relative phase between two amplitudes equal to +1 is considered here.
\begin{eqnarray}
&\frac{1}{2}(e_{H0}^2+f_{H0}^2)\xrightarrow{BS}\frac{1}{2}(a_{H0}^2+b_{H0}^2) \nonumber \\
&\xrightarrow{M} \frac{1}{4}(a_{H0}-b_{H0})^2+\frac{1}{4}(c_{H0}+d_{H0})^2 \nonumber \\
&\xrightarrow{BS} \frac{1}{2}(e_{H0}^2+f_{H1}^2).
\end{eqnarray}
The input state is redistributed in a sense that one amplitude is on the right side of the system and the other amplitude is on the left side while maintaining the original structure of the state.

HOM state with relative phase between two amplitudes equal to -1 is considered here.
\vspace{-5px}
\begin{eqnarray}
&\frac{1}{2}(e_{H0}^2-f_{H0}^2) \xrightarrow{BS} -a_{H0}b_{H0} \nonumber \\
&\xrightarrow{M} \frac{1}{4}(a_{H0}-b_{H0})^2-\frac{1}{4}(c_{H0}+d_{H0})^2 \nonumber \\
&\xrightarrow{BS} \frac{1}{2}(e_{H0}^2-e_{H1}^2).
\end{eqnarray}
In both cases, the output state is identical to the input state except for the spatial modes.
\vspace{-10px}
\subsubsection{Distinguishable photons}
\vspace{-10px}

Now, we examine the case of distinguishable two photon input.
The procedure is identical to the the previous case.
We begin with two distinguishable photons at each modes without superposition.
\begin{eqnarray}
&e_{H0}f_{V0} \xrightarrow{BS} \frac{1}{2}(a_{H0}-b_{H0})(a_{V0}+b_{V0})\nonumber \\
&\xrightarrow{M} -\frac{1}{2}(a_{H0}-b_{H0})(c_{V0}+d_{V0})\nonumber\\
&\xrightarrow{BS} - e_{H0} e_{V1}.
\end{eqnarray}

We examine the case of HOM states.
\begin{eqnarray}
&\frac{1}{2}(e_{H0}^2 \pm f_{V0}^2) \xrightarrow{BS}\frac{1}{4}(a_{H0}-b_{H0})^2\pm\frac{1}{4}(a_{V0}+b_{H0})^2\nonumber\\
&\xrightarrow{M}\frac{1}{4}(a_{H0}-b_{H0})^2\pm\frac{1}{4}(c_{V0}+d_{H0})^2\nonumber\\
&\xrightarrow{BS} \frac{1}{2}(e_{H0}^2 \pm e_{V1}^2).
\end{eqnarray}

\begin{figure}[htp]
\subfloat[]{%
  \includegraphics[clip,width=\columnwidth]{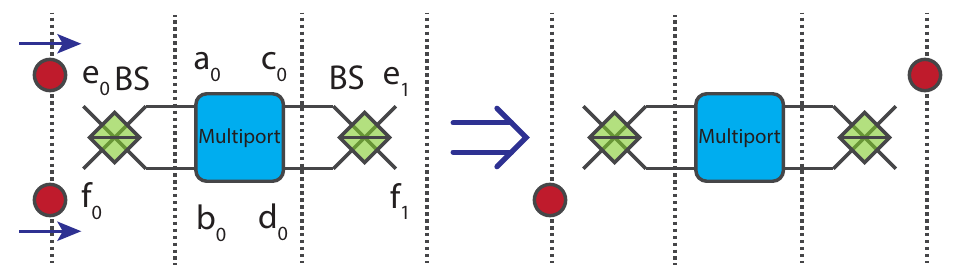}%
}
\hspace{-10px}
\subfloat[]{%
  \includegraphics[clip,width=\columnwidth]{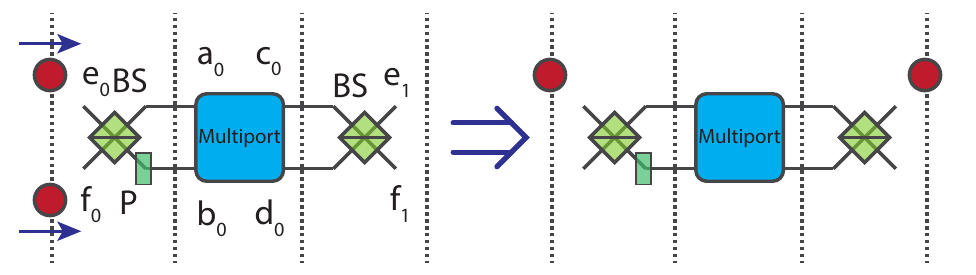}%
}
\hspace{-10px}
\subfloat[]{%
  \includegraphics[clip,width=\columnwidth]{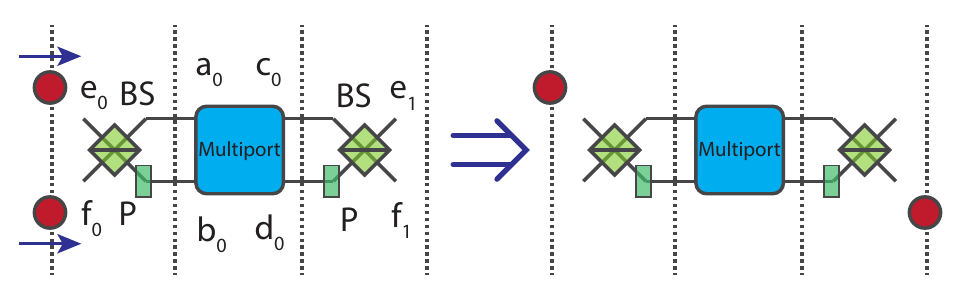}%
}
\hspace{-10px}
\subfloat[]{%
  \includegraphics[clip,width=\columnwidth]{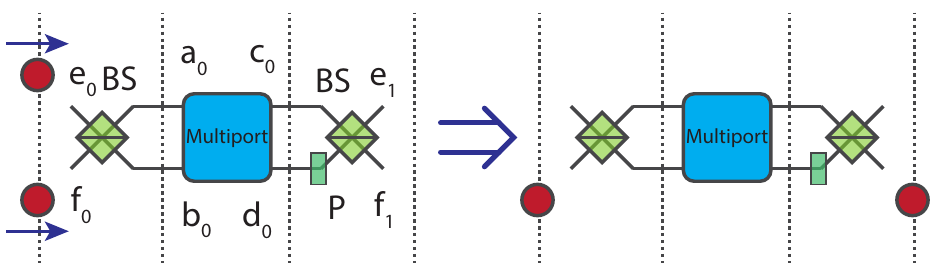}%
}

\caption{Quantum state redistribution with control of propagation direction. We performed the same analysis as the higher dimensional HOM effect with direction control. By introducing phase shifters in the system before beam splitters, we can change the exit direction of the amplitudes. The starting state is $e_0f_0$. The first beam splitter transforms the input state, then they enter the multiport device. The multiport transformed state goes through beam splitters on the right and left side. The final outcome has the same form as the input state.}
\label{fig:control_without_circ}
\end{figure}

The control of exit location can be performed as well in this scheme by introducing phase shifters in the system as indicated in Fig. \ref{fig:control_without_circ}.
This procedure does not destroy the redistribution property.
There are four potential spatial modes and by switching the phase shift before beam splitters, the direction of propagation switches.
The combinations are, (a): $(e_0,f_0)\rightarrow(e_0,e_1)$, (b): $(e_0,f_0)\rightarrow(e_0,f_1)$, (c): $(e_0,f_0)\rightarrow(f_0,e_1)$, and (d): $(e_0,f_0)\rightarrow(f_0,f_1)$.

The result from the system with circulators is summarized in Table. \ref{tab:table_1}, the system without them is in Table. \ref{tab:table_2}.
In the case of indistinguishable photons, the results are cyclic in a sense that all three states can be produced by using the other system.
However, there is a significant difference when distinguishable photons are considered.

\begin{table}
\begin{tabular}{P{0.4\linewidth}P{0.6\linewidth}}
\hline \hline
\noalign{\vskip 1ex}
\multicolumn{2}{c}{State transformation with circulators}\\
\noalign{\vskip 1ex}

\hline \hline

\noalign{\vskip 1ex}
Indistinguishable photons & $a_{H0}b_{H0}  \rightarrow -\frac{1}{2}(e_{H0}^2 - e_{H1}^2)$ \\
\noalign{\vskip 1ex}

\hline

\noalign{\vskip 1ex}
HOM pair with +1 relative phase & $\frac{1}{2}(a_{H0}^2 + b_{H0}^2) \rightarrow \frac{1}{2}(e_{H0}^2 + e_{H1}^2)$\\
\noalign{\vskip 1ex}

\hline

\noalign{\vskip 1ex}
HOM pair with $-1$ relative phase & $\frac{1}{2}(a_{H0}^2 - b_{H0}^2)  \rightarrow -e_{H0}e_{H1}$  \\
\noalign{\vskip 1ex}

\hline

\noalign{\vskip 1ex}
Distinguishable photons & $a_{H0}b_{V0} \rightarrow -\frac{1}{2}(e_{H0}-e_{H1})(e_{V0}+e_{V1})$ \\
\noalign{\vskip 1ex}

\hline

\noalign{\vskip 1ex}
Distinguishable HOM pair & $\frac{1}{2}(a_{H0}^2 \pm b_{V0}^2) \rightarrow \frac{1}{4}\{(e_{H0}-e_{H1})^2\pm(e_{V0}-e_{V1})^2\}$  \\
\noalign{\vskip 1ex}

\hline \hline

\end{tabular}
\caption{State transformations in a system with circulators.
The first three states deal with indistinguishable photons by giving them the same polarization.
A state consisting of two single photons will become an HOM state.
We analyzed HOM states as an initial state, and they become either the HOM state or a two single-photon state.
Distinguishable photons are also analyzed by introducing orthogonal polarizations.
The output states become coupled states meaning the original states are not preserved.
} \label{tab:table_1}
\end{table}

\begin{table}
\begin{tabular}{P{0.4\linewidth}P{0.6\linewidth}}
\hline \hline
\noalign{\vskip 1ex}
\multicolumn{2}{c}{State transformation without circulators} \\
\noalign{\vskip 1ex}

\hline \hline

\noalign{\vskip 1ex}
Indistinguishable photons & $e_{H0}f_{H0}  \rightarrow -e_{H0}e_{H1}$ \\
\noalign{\vskip 1ex}

\hline

\noalign{\vskip 1ex}
HOM pair with +1 relative phase & $\frac{1}{2}(e_{H0}^2 + f_{H0}^2) \rightarrow \frac{1}{2}(e_{H0}^2 + e_{H1}^2)$  \\
\noalign{\vskip 1ex}

\hline

\noalign{\vskip 1ex}
HOM pair with $-1$ relative phase & $\frac{1}{2}(e_{H0}^2 - f_{H0}^2) \rightarrow \frac{1}{2}(e_{H0}^2 - f_{H1}^2)$   \\
\noalign{\vskip 1ex}

\hline

\noalign{\vskip 1ex}
Distinguishable photons & $e_{H0}f_{V0} \rightarrow -e_{H0}e_{V1}$  \\
\noalign{\vskip 1ex}

\hline

\noalign{\vskip 1ex}
Distinguishable HOM pair & $\frac{1}{2}(e_{H0}^2 \pm f_{V0}^2) \rightarrow \frac{1}{2}(e_{H0}^2 \pm f_{V1}^2)$ \\
\noalign{\vskip 1ex}

\hline \hline

\end{tabular}
\caption{State transformations in a system without circulators.
The structure of the table is the same as the Table. I.
The first three states deal with indistinguishable photons by giving them the same polarization.
The last two states handle distinguishable photons.
The output states preserve the same form as the input state.
We start the transformation from the system location 0, then the transformed states are redistributed between location 0 and location 1.
The result shows coherent transportation of input states.} \label{tab:table_2}
\end{table}

\vspace{-10px}
\subsection{Delayed state redistribution}
\vspace{-10px}
\begin{figure*}[htp]
\centering
\subfloat[][State redistribution without reflection]{\includegraphics[clip,width=\columnwidth]{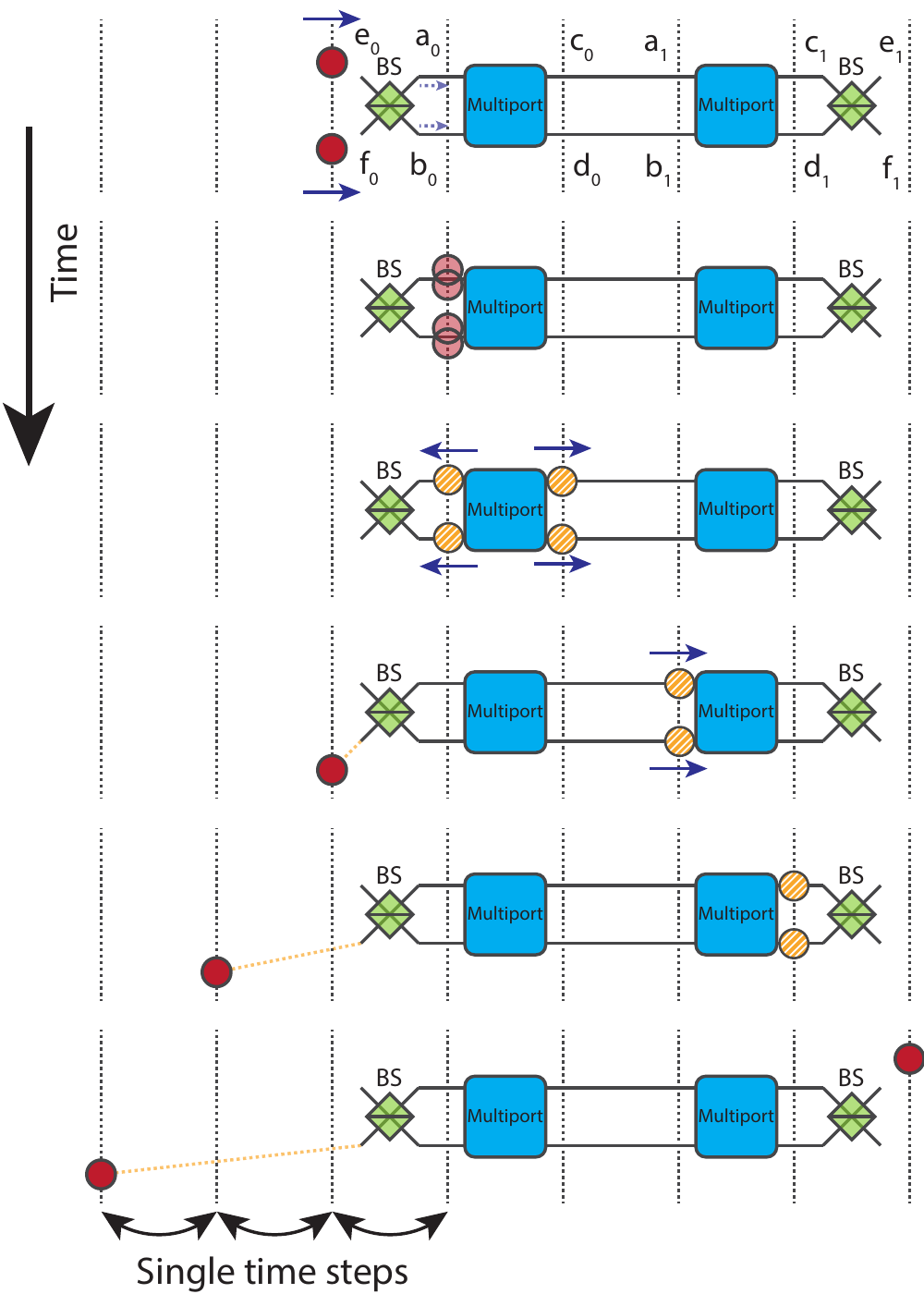}\label{}}
\subfloat[][State redistribution with reflection]{\includegraphics[clip,width=\columnwidth]{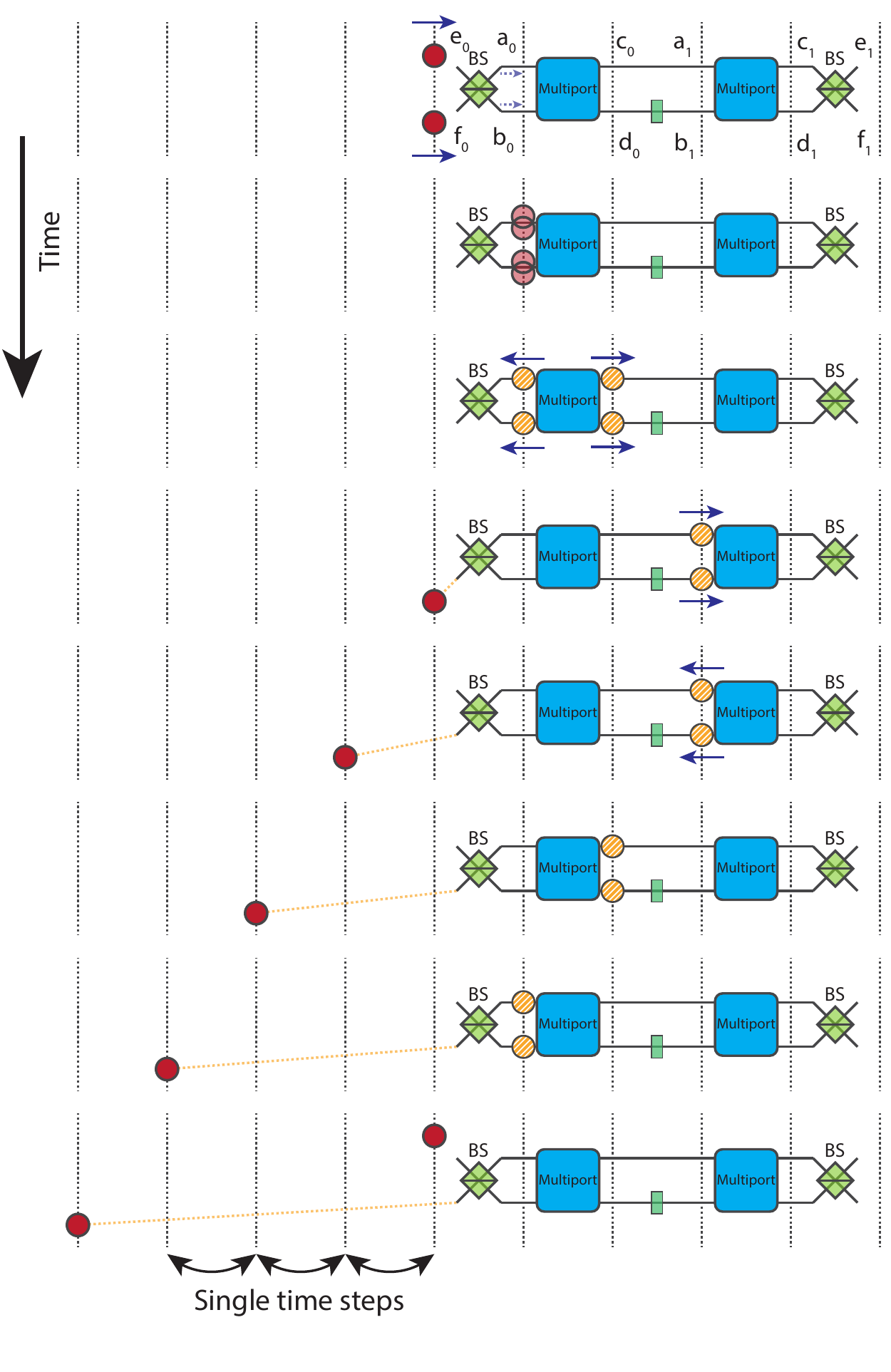}\label{}}
\caption{Delayed state redistribution.
The two-photon amplitude transformation progresses in time from top to bottom.
The distance traveled in a single time step is indicated by vertical dashed lines. The total photon numbers are two in the system through out the propagation. At the first step for both cases, the input two-photon state is transformed by the BS.
The transformed state becomes the HOM state, and it is indicated as red transparent overlapped circles occupying both modes.
The initial and the final transformed state are indicated using solid red circles, and intermediate states are indicated in striped yellow circles.
(a) Two multiports and beam splitters {\it without} phase shifters between the multiports.
The HOM state enters the multiport and transformed taking the form of $-\frac{1}{2}(a_{0}-b_{0})(c_{0}+d_{0})$.
The amplitudes are coupled, however, they propagate without changing its amplitude. After several steps, the amplitudes occupying two rails converges to a single mode state after transformation by beam splitters.
The final state has the same form as the input state.
(b) Two multiports and beam splitters {\it with} a phase shifter P set at $\pi$ between multiports.
When the P is present, the right-moving coupled amplitude gains a relative phase between modes $a_1$ and $b_1$.
Reflection occurs at the multiport when the relative phase between the two is $\pi$.
Therefore, the transformed amplitude reflects upon a second multiport encounter, going back to the original state with opposite propagation direction.
Reflection does not occur on this transformed coupled left-moving amplitude, therefore it continues to propagate leftward.
The original left-moving amplitude becomes available for detection earlier than the transformed left-moving amplitude.}
     \label{fig:delayed_state_dist}
\end{figure*}

We introduce the temporal delay effect as the higher dimensional HOM case by introducing a phase shifter between two multiports.
\vspace{-10px}
\subsubsection{Without reflection}
\vspace{-10px}
When there is no phase shifter between the two multiports, the result is identical to the system with a single multiport from the previous section.
The state transformation and propagation is provided schematically in Fig. \ref{fig:delayed_state_dist} (a).
The photons are initially sent from the left side of the BS.
The correlated photons are transformed to HOM state through the BS.
\vspace{-10px}
\begin{eqnarray}
&{e_{0}f_{0}}_R \xrightarrow{BS} \frac{1}{2}(a_{0}^2-b_{0}^2)_R \nonumber\\
&\xrightarrow{M}-\frac{1}{2}(a_{0}-b_{0})_L(c_{0}+d_{0})_R.
\end{eqnarray}
The HOM state is transformed by the multiport device.
This state is in a coupled state because right moving and left moving amplitudes are not separated.
We propagate this state through the BS on the left and translate the amplitudes moving to the right.
\begin{eqnarray}
&\xrightarrow{BS,T}-\frac{1}{\sqrt{2}}f_{0L}(a_{1}+b_{1})_R \xrightarrow{M} -\frac{1}{\sqrt{2}}f_{0L}(c_{1}+d_{1})_R\nonumber\\
&\xrightarrow{BS} -f_{0T_0L}e_{1T_1R}
\end{eqnarray}
The left moving amplitude is transformed by the left BS while right moving amplitude propagates to the second multiport device.
We introduced temporal difference between the right moving and the left moving photons.

\vspace{-15px}
\subsubsection{With reflection}
\vspace{-10px}
Reflection of amplitudes are introduced when there is a phase shifter between two multiport devices as indicated in fig. \ref{fig:delayed_state_dist} (b).
\begin{eqnarray}
&{e_{0}f_{0}}_R \xrightarrow{BS} \frac{1}{2}(a_{0}^2-b_{0}^2)_R
\xrightarrow{M} -\frac{1}{2}(a_{0}-b_{0})_L(c_{0}+d_{0})_R \nonumber \\
&\xrightarrow{BS,T+P}-\frac{1}{\sqrt{2}}f_{0L}(a_{1}-b_{1})_R
\end{eqnarray}
The right moving amplitude gains relative phase between upper and lower rails, and this relative phase allows the amplitude to get reflected upon multiport encounter.
\begin{eqnarray}
&\xrightarrow{M} -\frac{1}{\sqrt{2}}f_{0L}(a_{1}-b_{1})_L \xrightarrow{T+P} -\frac{1}{\sqrt{2}}f_{0L}(c_{0}+d_{0})_L\nonumber\\
&\xrightarrow{M} -\frac{1}{\sqrt{2}}f_{0L}(a_{0}+b_{0})_L \xrightarrow{BS} -f_{0T_{0}L} e_{0T_{2}L}
\end{eqnarray}
The input photons do not have any delays between the two at the beginning. The delay  $\Delta T = T_2 - T_0$ is introduced from the reflection in the system.

\vspace{-10pt}
\section{Conclusion}
\vspace{-10px}
We demonstrated higher dimensional quantum state manipulation such as the HOM effect and state redistribution by applying linear-optical four-ports realizing four-dimensional Grover matrix accompanied by beam splitters and phase shifters.
Identical photons are sent into two of the four input-output ports and split into right-moving and left-moving amplitudes, with no cross terms to observe the HOM effect.
This absolute separation of propagation direction without mixing of right-moving and left-moving amplitudes insures the photons remain clustered as they propagate through the system.
Variable phase shifts in the system allow the HOM photon pairs to switch between four spatial output destinations, which can increase information capacity.
Time delays between emerging parts of the clustered two-photon state illustrating “delayed” HOM effect can be engineered using two multiports.
In addition, depending on the phase shifter position, the propagation direction can be reversed so that the right moving amplitude can get reflected at the second multiport, resulting in HOM pairs always leaving only from the left side of the system and with a particular time-bin delay.
The same situations have been investigated in a system without circulators. This system allows to redistribute the input state between the right and the left side of the system without changing amplitudes.
The HOM effect and clustered photon pairs are widely used in quantum information science. The approach introduced here adds extra degrees of freedom, and paves the way for new applications that require control over the spatial and temporal modes of the HOM amplitudes as they move through one- and two-dimensional networks.
We have demonstrated two photon amplitude control in both spatial and temporal modes. This two photon system can be extended to multiphoton input states, and manipulation of more complex entangled states would be the next milestones to be achieved.

\vspace{-15pt}
\section*{Appendix}
\appendix*

In this Appendix we briefly comment on the structure of the transformations carried out by the apparatus with circulators (Fig. \ref{fig:multiport_circ}) and without (Fig. \ref{fig:multiport_no_circ}). We assume all the input is coming from the left side, and we look at the unitary matrix that evolves the input for the amount of time required either to transmit out the right side or to reflect back out the left side. This will be represented by a six-by-six matrix: the first two rows and columns represent the external lines on the left side, the last two represent the external lines on the right side. The middle two rows and columns represent the lines between the two multiports. Some of the input will remain in this central region initially, so these interior lines must be explicitly kept in  order to maintain unitarity. 

Let $U$ represent the single-photon evolution matrix \emph{with} circulators, as in Fig.  \ref{fig:multiport_circ}, while $V$ represents the same for the case \emph{without} circulators, Fig.  \ref{fig:multiport_no_circ}. The difference in the two cases is simply that for $U$ the ingoing amplitudes bypass the first beam splitter (but any left-moving output amplitude will encounter it on the way out), while for $V$ the amplitude on the left will see the beam splitter both on the way in and the way out.

The evolution matrices can be written as
\vspace{-5pt}
\begin{eqnarray}
U &=& (BS_1P_1)(BS_2P_2M_2)M_1P_1 \label{Uev} \\
V &=& (BS_1P_1)(BS_2P_2M_2)M_1P_1BS_1\label{Vev}\nonumber\\
&=&  U\cdot BS_1 .
\end{eqnarray}

Here $BS_1$ and $BS_2$ represent the left and right beam splitters, $P_1$ and $P_2$ represent the left and right phase shifters (which can each be set to $0$ or $\pi$), and $M_1$ and $M_2$ represent the left and right multiports. In the expression for $U$, the terms in the first set of parentheses act on the amplitude that reflects backwards from $M_1$, while the terms in the second set act on the portion of the amplitude transmitted at $M_1$.

We attach subscripts (hence $U_j$ and $V_j$) to indicate the various cases discussed in the main text:
$j=0$ indicates that all the phases are set to $0$, $j=2$ and $j=6$ represent the cases the phases are $\pi$ in the lower left and lower right lines respectively, with all other phases zero. $j=26$ indicates the case with $\pi$ at both lower lines, left and right.
Then $U_j$ and $V_j$ have the general forms:
\vspace{-5px}
\begin{equation}
U_j=
\begin{pmatrix}
A_j & \rvline &
B_j & \rvline &
\begin{matrix}
\; 0 & 0  \\
\; 0 & 0
\end{matrix}\\ 
\hline
\begin{matrix}
0 & 0  \\
0 & 0
\end{matrix} & \rvline &
\begin{matrix}
\frac{1}{2} & -\frac{1}{2}  \\
-\frac{1}{2} & \frac{1}{2}
\end{matrix} & \rvline &
\begin{matrix}
\frac{1}{2} & \frac{1}{2}  \\
\frac{1}{2} & \frac{1}{2}
\end{matrix}\\ 
\hline
C_j & \rvline &
\begin{matrix}
 0 & \; 0  \\
 0 & \; 0
\end{matrix} & \rvline &
D_j &
\end{pmatrix} ,\label{Umatrix}
\end{equation}
\begin{equation}
V_j=
\begin{pmatrix}
  E_j & \rvline &
  B_j & \rvline &
   \begin{matrix}
  \; 0 & 0  \\
  \; 0 & 0
  \end{matrix}\\ 
  \hline
\begin{matrix}
  0 & 0  \\
  0 & 0
  \end{matrix} & \rvline &
   \begin{matrix}
  \frac{1}{2} & -\frac{1}{2}  \\
  -\frac{1}{2} & \frac{1}{2}
  \end{matrix} & \rvline &
   \begin{matrix}
  \frac{1}{2} & \frac{1}{2}  \\
  \frac{1}{2} & \frac{1}{2}
   \end{matrix}\\ 
   \hline
  F_j & \rvline &
 \begin{matrix}
  0 & \; 0  \\
  0 & \; 0
  \end{matrix} & \rvline &
   D_j &
\end{pmatrix}.\label{Vmatrix}
\end{equation}

Depending on where the phase shifts are located, the $A$, $B$, $C$, $D$ matrices are always of one of the forms
\begin{equation}
\pm\left( \begin{array}{cc} 0 & 0  \\ \frac{1}{\sqrt{2}}&\pm\frac{1}{\sqrt{2}}\end{array}\right) \qquad  \mbox{or} \qquad \pm \left( \begin{array}{cc}  \frac{1}{\sqrt{2}}&\pm \frac{1}{\sqrt{2}} \\ 0 & 0 \end{array}\right) .
\end{equation} 
The $E$ and $F$ submatrices always have one entry equal to $\pm 1$ and the other three entries vanishing, with the position of the nonzero entry depending on the locations of the phase shifts in the apparatus.

The corresponding matrices for two-photon transformations would be $36\times 36$, and can be found by taking tensor products of matrices of the forms above. Because of the different locations of the plus and minus signs in the $A$ and $B$ submatrices, different phase shifter positions lead to different combinations of constructive and destructive interference, and therefore the output clusters in different pairs of exit ports, as seen in the main text.

The expressions given in Eqs. \ref{Umatrix} and \ref{Vmatrix} are lowest order approximations to the true transformation matrix. Higher order terms occur due to the possibility of the photon amplitudes bouncing back and forth several times between the two multiports. These higher order terms are obtained by replacing $M_2M_1$ in Eqs. \ref{Uev} and \ref{Vev} by higher iterates $\left( M_2M_1\right)^n$, with $n>1$. For fixed input state, the amplitudes at the outputs of these higher iterations will decay as $n$ increases.

\vspace{10pt}

This research was supported by the National Science Foundation EFRI-ACQUIRE Grant No. ECCS-1640968, AFOSR Grant No.FA9550-18-1-0056, and DOD/ARL Grant No.W911NF-20-2-0127.

\vspace{10pt}

\bibliographystyle{apsrev4-2}
\bibliography{HOM_bib}


\end{document}